# On the Spectral Slopes of Hard X-ray Emission from Black Hole Candidates

Ken EBISAWA,[*] Lev TITARCHUK[†]

*code 660.2, NASA/GSFC, Greenbelt, MD 20771, USA*

and

Sandip K. CHAKRABARTI

*Tata Institute of Fundamental Research, Bombay, 400005 India*

December 3, 1995

**Abstract**

Most black hole candidates exhibit characteristic power-law like hard X-ray emission above $\sim$ 10 keV. In the *high state*, in which 2 – 10 keV luminosity is relatively high, the energy index of the hard X-ray emission is usually greater than 1 — typically $\sim$ 1.5. On the other hand, in the *low state*, the hard X-ray energy index is 0.3 – 0.9. In this paper, we suggest that this difference of the hard X-ray spectral slopes may be due to two different Comptonization mechanisms. We propose that, in the high state, the hard component is governed by the Comptonization due to the bulk motion of the almost freely falling (convergent accretion) flow close to the black hole, rather than thermal Comptonization. The spectral slope of the hard component is insensitive to the disk accretion rate governing the soft component, hence is nearly invariant in spite of the soft component variations. The power-law component due to the bulk motion Comptonization has a sharp cut-off at around the electron rest mass energy, which is consistent with high energy observations of the high state. In the low state, the spectrum is formed due to thermal Comptonization of the low-frequency disk radiation by a sub-Keplerian component (possibly undergoing a centrifugally-supported shock) which is originated from the Keplerian disk. In the limit of low disk accretion rate, the power law index is uniquely determined by the mass accretion rate of the sub-Keplerian component.

[*]Universities Space Research Association

[†]George Mason University/CSI

# 1 Introduction

Galactic black hole candidates have been known to show interesting luminosity/spectral properties. Most black hole candidates are categorized as either in the high state (soft state) or in the low state (hard state). Usually, in the high state, the 2 − 10 keV luminosity is relatively higher than in the other state, and the energy spectra can be represented with a soft, thermal component ($T \lesssim 1$ keV) and a power-law like hard component above ∼ 10 keV. On the other hand, in the low state, the energy spectrum in 2 − 50 keV is roughly represented by a single power-law function with a smooth turnover at higher energies. Occasionally, the same object shows transitions between the two states.

In this paper, we emphasize the difference of the spectral slopes of the hard components between the two states and explain their origins. In Table 1, we show examples of the observed spectral slopes above 10 keV in the both states (throughout this paper, the spectral slope, $\alpha$, denotes the *energy* index, with which the energy emitted per frequency is proportional to $E^{-\alpha}$. The *photon index* is $\alpha + 1$). We see that the energy spectral index is $\geq 1$ (typically, $1.3 - 1.8$) in the high state and $\leq 1$ (typically, $0.3 - 0.8$) in the low state. We show that this dichotomy in the energy spectral index is understood by considering relative importance between the thermal Comptonization (Sunyaev & Titarchuk 1980, 1985; Titarchuk 1994a; Titarchuk & Lybarskij 1995) and Comptonization due to bulk motion (Blandford & Payne 1981; Hirotani, Hanawa and Kawai 1990; Hanawa 1991; Titarchuk, Mastichiadis & Kylafis 1995, hereafter TMK95). We propose that the steep spectral slope in the high state is due to the bulk motion, and the flatter slope in the low state is due to the thermal Comptonization.

A successful explanation of this long-recognized puzzle hinges on understanding the very nature of the accretion disk around a black hole. In the most general case, an accretion disk has both the Keplerian and non-Keplerian components (e.g., Chakrabarti 1990; Chakrabarti & Molteni 1995; Chakrabarti, Titarchuk, Kazanas & Ebisawa 1995; Chakrabarti & Titarchuk 1995, hereafter CT95). From the complete solution (Chakrabarti 1990; Chakrabarti 1996) of viscous transonic equations (e.g. Matsumoto et al. 1984; Abramowicz et al. 1988; Chen & Taam 1993), it is demonstrated that for a given cooling process, there is a critical viscosity parameter, $\alpha_c$: an originally Keplerian disk with viscosity less than $\alpha_c$ becomes sub-Keplerian very far away from the black hole (where the Mach number of the flow is almost 0), and the accreting matter passes through the outer sonic point (and possibly a shock after that) before entering the black hole. However, Keplerian disk with viscosity higher than the critical value deviates from Keplerian only very close to the black hole. Thus, when a disk viscosity is vertically falling off, for a given cooling process, the highly viscous Keplerian disk (Shakura & Sunyaev 1973; Novikov & Thorne 1973) resides on the equatorial plane, while the sub-Keplerian component resides above and below it. The sub-Keplerian component, having the angular momentum $l$, can form a standing shock wave (e.g., Chakrabarti 1990) which heats up the disk to a virial temperature of $T_p \sim 10^{11}$K, on hitting its centrifugal barrier. This takes place at the radius $r_{sh}$ where $l^2_{Kep}(r_{sh}) \sim l^2$, for $l$ corresponding to the marginally bound orbit; in the Schwarzschild geometry $r_{sh} \approx 8 \sim 10\, r_g$ ($r_g$ is the Schwarzschild radius $2GM/c^2$), and roughly half as distant in the Kerr geometry (CT95).

In the soft state, when the accretion rate is higher, the soft photons from the Keplerian disk completely cools the post-shock region due to thermal Comptonization. The resulting cooler converging inflow, as it rushes towards the black hole, traps the soft-photons within the trapping radius of $r_t \sim \dot{M}_{conv}/\dot{M}_E$, where $\dot{M}_{conv}$ is the net accretion rate and $\dot{M}_E$ is the Eddington accretion rate, and transfers its momentum to the soft-photons to produce a weak power-law component with an energy spectral slope of $\alpha \sim 1.5$. On the other hand in the hard state, the accretion rate of the Keplerian disk is not high enough to cool the Comptonization process in the post-shock region, and the hotter post-shock region forms a strong power-law component with $\alpha < 1.0$. Thus, in the soft state, the energy spectrum contains optically thick disk spectrum and a weak power-law component due to the bulk motion Comptonization, whereas in the hard state, the spectrum is produced due to thermal Comptonization. An explanation of state transitions due to variation of soft photon numbers is also presented in Mineshige, Kusunose and Matsumoto (1995). Note that in our model, the high/low state transition is due to variation of the dimensionless mass accretion rate normalized by the Eddington mass accretion rate. Therefore, for sources with different masses, different spectral states occur for the same luminosity (see, e.g., Tanaka 1989; Miyamoto et al. 1995). Below we show the two distinct spectral behaviors are naturally derived on the above considerations.

## 2 Computation of Spectral Indices

### 2.1 General description

As was shown by Zel'dovich (see, review by Pozdnyakov, Sobol & Sunyaev 1983) the spectral index of the up-scattered spectrum can be understood in terms of the scattering probability $p$ and the photon energy gain $\eta$ per scattering. The intensity of the out-coming photons $I_k$ which undergo $k$-th scatterings in the plasma cloud is proportional to $p^k$, i.e.

$$I_k \propto p^k \tag{1}$$

On the other hand, the photon with the initial energy $E_0$ gains energy

$$E = E_0(1+\eta)^k \tag{2}$$

after $k$-th scattering. Substitution of $k$ expressed through $\eta$ (see Eq. 2) to equation (1) makes the power-law dependence of the intensity $I_k = I_\nu \propto (E/E_0)^{-\alpha}$ on energy $E$ with the spectral index

$$\alpha = \frac{\ln(1/p)}{\ln(1+\eta)}. \tag{3}$$

The scattering probability $p$ is determined by the number of scattering $N_{sc}$ underwent by the emergent photon in the plasma cloud. Because of the probability of the escape after the mean number of scatterings is, by definition, equals to one, i.e., $N_{sc}(1-p) = 1$ we can rewrite equation (3) as follows:

$$\alpha = \frac{\ln(1 - 1/N_{sc})^{-1}}{\ln(1 + \eta)}. \tag{4}$$

The exact spectral index formula in the case of the pure thermal Comptonization as a function of the plasma cloud temperature and optical depth is presented in Titarchuk & Lyubarskij (1995).

Below we shall show how this approach can be applied for explanation of the spectral properties in the low and high states of black holes.

## 2.2 In High State

In what follows, we use the Schwarzschild radius $r_g$ to be the unit of distance and the Eddington accretion rate $\dot{M}_E \equiv 4\pi G M m_p/\sigma_T c$ to be the unit of the accretion rate, where $M$, $G$, $m_p$ and $\sigma_T$ are the mass of the black hole, the gravitational constant mass of the proton, and the Thomson cross-section, respectively. Accreting matter moving with velocity $v(r) \sim r^{-1/2}$ produces a density distribution of $n(r) \propto \dot{m} \, r^{-3/2}$. If Comptonization is the dominant mechanism for cooling, the equation governing the electron temperature is the following (CT95);

$$\frac{1}{T_e}\frac{dT_e}{dr} + \frac{1}{r} - C_{Comp}r^{1/2} \sim 0, \tag{5}$$

where $C_{Comp}$ is roughly a spatially-constant but otherwise monotonically increasing with accretion rate (optical depth). The solution of the above equation is

$$T_e \propto \frac{r_{sh}}{r}e^{C_{Comp}(r^{3/2}-r_{sh}^{3/2})}. \tag{6}$$

This shows that if $C_{Comp} > r_{sh}^{-3/2}$, i.e., accretion rate is high enough, the cooling due to Comptonization overcomes compressional (geometric) heating and $T_e$ drops as the flow approaches the black hole and the convergent inflow regime begins. In this case, The resulting Thomson opacity $\tau_T \sim \dot{m}r^{-1/2}$ is high enough to make the diffusion time scale $r\tau_T r_g/c$ comparable to the advection time scale $r\, r_g/v(r)$. Below the trapping radius, $r < r_{tr} \sim \dot{m}$ (Rees 1978; Begelman 1979), the advection dominates and the photons are bound to be advected along with the flow. Only a fraction of those photons which did not reach the black hole horizon diffuses outward producing a power-law hard component with a sharp cutoff after being Comptonized by the bulk motion.

Let us investigate the relative importance of the Comptonization due to bulk motion (convergent flow) and that due to thermal motion. The energy gain of a photon per scattering by thermal Comptonization $<\Delta E_{comp}>$ is proportional to $(v/c)^2$, i.e.,

$$<\Delta E_{compt}> = E\frac{4kT_e - E}{m_e c^2}$$

(e.g., Rybicki & Lightman 1979).

On the other hand, the energy gain in the presence of the bulk motion, $<\Delta E_{cf}>$ is proportional to $v/c$, (e.g. Blandford & Payne 1981; Lyubarskij & Sunyaev 1982; TMK95), i.e.,

$$<\Delta E_{cf}> = E\frac{d(v/c)}{d\tau_T} = \frac{1}{\dot{m}}.$$

Thus, we have

$$\frac{<\Delta E_{compt}>}{<\Delta E_{cf}>} = \frac{(4kT_e - E)\dot{m}}{m_e c^2} < \frac{4}{\delta},$$

where $\delta = 51.1 \times T_{10}^{-1}\dot{m}^{-1}$ and $T_{10} = T_e/10\text{keV}$.

Hence, the thermal Comptonization is negligible if $\dot{m}T_{10} < 13$. CT95 shows that $T_e \lesssim 10$ keV in the soft state, hence the bulk motion Comptonization must be more important than thermal Comptonization.

The origin of the power-law and the cut-off could be understood in the following way: as shown above, the change in energy of the soft photon due to collision with accretion matter of the bulk velocity $v(r)$ having the electron temperature $T_e$ is given by,

$$\eta = \frac{<\Delta E>}{E}\Big|_{Dopp} \sim \dot{m}^{-1} + \frac{4kT_e}{m_e c^2}. \tag{7a}$$

At the same time, recoil effects change the energy by:

$$\frac{<\Delta E>}{E}\Big|_{rec} \sim -\frac{E}{m_e c^2}. \tag{7b}$$

When the photon energy $E$ is much lower than the electron rest mass energy, the recoil effect is negligible, resulting in a pure power-law spectrum with the index $\alpha$ given by equation (3). At higher energies, these two effects are comparable, and the cutoff is seen at $E/m_e c^2 \sim \dot{m}^{-1} + 4kT_e/m_e c^2 \sim \dot{m}^{-1}$ (since $\dot{m}^{-1} \gg 4kT_e/m_e c^2$ in the soft state).

In order to understand the origin of the slope of $\sim 1.5$, one requires to solve an equation of the spectral energy flux $F(r, E)$ with the boundary conditions that $F \to 0$ at $r \to \infty$ (the outer boundary of the converging inflow) and $F = -\frac{1}{2}nx^3$ at the inner boundary on the horizon. Here, $n$ is the photon occupation number and $x = E/kT_e$, therefore $nx^3$ is the average intensity of photons. The factor $1/2$ is introduced because a black hole absorbs all the soft photons in $2\pi$ solid angle (here we have ignored the effects of the gravitational bending of light). Exact solution of this problem (TMK95) indicates that the average number of scattering $N_{\text{sc}}$ that emergent photons undergo in order to form the power-law hard tail is given by $N_{\text{sc}} \approx 2\dot{m}/3$. The number of scattering $N_{\text{sc}}$ is proportional to the total optical depth $\tau_0$, i.e $N_{\text{sc}} \approx 2\dot{m}/3 = 2\tau_0/3$. This is to be contrasted with the general notion that $N_{\text{sc}} \propto \tau_0^2$ (Sunyaev and Titarchuk 1980), since observations select only those photons which undergo $\pi$-angle scattering, i.e., only those photons which scatter near radial direction gain energy from the converging inflow, ultimately forming the hard component that is observed.

Now, from eq. 2, after $k$-th scattering the energy of photon becomes,

$$E = E_0(1 + \dot{m}^{-1})^k, \tag{8}$$

which gives,

$$k = \frac{\ln(E/E_0)}{\ln(1 + \dot{m}^{-1})}. \tag{9}$$

The intensity of radiation from such a scattering (see Eq. 4),

$$I \propto (1 - N_{sc}^{-1})^k = (1 - 1.5/\dot{m})^k \sim E^{-1.5} \quad \text{for} \quad \dot{m} \gg 1. \tag{10}$$

This slope of 1.5 which roughly corresponds to that seen in most of the objects in the high state (Table 1) is insensitive to the accretion rate when $\dot{m}$ is high enough. In the above derivation, we assumed spherical converging inflow geometry. Agreement with observations does therefore indicate that the geometry of the flow close to the black hole is roughly spherical and sub-Keplerian. Occasionally, in the soft state, the hard tail is not observed at all. In this case, we believe that though the disk accretion rate is high enough to cool down the post-shock region, it is not sufficient to form the bulk motion power-law spectrum (CT95).

## 2.3 In Low State

In the low state, the hard spectrum is formed as a result of Comptonization of the low-frequency disk radiation in the hot post-shock region (Chakrabarti, Titarchuk, Kazanas & Ebisawa 1995; CT95) located close to the black hole ($5 - 10\ r_g$) depending on the black hole angular momentum. In this case, disk accretion rate is very low and therefore the supply of soft photons are low as well. When both the cooling due to Comptonization and heating due to the Coulomb process are important, the electron temperature roughly becomes $T_e = (m_p/m_e)^{1/2}\ T_p$. Average temperature of the proton in the post-shock region is (at $r \sim 10$) $T_p \sim 1.5 \times 10^{11}$K, and thus $T_e \sim 3.3 \times 10^9$K. The total energy release in the Comptonization process in the post-shock region with the optical depth $\tau_0$ is (e.g. Zel'dovich & Shakura 1969),

$$\frac{Q}{\tau_0} \sim 15\ \alpha\ \varepsilon(\tau)\ T_e\ /f(T_e). \tag{11}$$

Here, $\tau$ is the current Thomson optical depth in the post shock region, $\alpha$ is the energy spectral index of the power law part of the Comptonization spectrum, and $\varepsilon(\tau)$ is the radiative energy density distribution over the shock region. The $f(T_e)$ is a function of $T_e$, and $f(T_e) \sim 1 + 2.5T_e/m_ec^2 \sim 2.5$. The $\varepsilon(\tau)$ is given by,

$$\varepsilon(\tau) = \frac{Q}{c}\{3^{1/2} + 3\tau_0[\tau/\tau_0 - 0.5(\tau/\tau_0)^2]\}. \tag{12}$$

This is $\sim 3Q/c$ for $\tau_0 \approx 1$ which is valid for one Eddington rate for the sub-Keplerian matter. Thus, from eq. (12), we get,

$$\alpha\ T_e \tau_0 \sim 1.67 \times 10^9 = \text{const.} \tag{13}$$

Since the spectral index is a function of the product $T_e\tau_0$ only (for $\tau_0 \leq 1$; e.g. TL95), eq. (13) gives $\alpha \sim$ constant. With the $\tau_0$ and $T_e$ computed above, we get, $\alpha \sim 0.5$, which is the typical observed index in the hard state. As long as the disk rate is much smaller compared to the sub-Keplerian rate (which determines $\tau_0$), the result is insensitive to the disk rate.

In Fig. 1, we present schematically the nature of the accretion flows which typically contains the Keplerian and sub-Keplerian components in varying degree depending upon the state of a black hole. Our scheme includes not only the soft state and the hard state, but also the quiescent state and the pre-outburst state which are appropriate only for black hole transients. In the soft state (Case c), the Keplerian disk accretion rate is high because the viscosity is high. The spectrum is dominated by the soft X-rays from the disk and the weak hard component due to the convergent inflow (dashed line). In the hard state (Case d), the disk accretion rate is low since the viscosity is small. In the quiescent state of black hole transients, the viscosity is so small that the Keplerian disk is formed only very far away from the black hole (Chakrabarti 1990, 1996), as shown in Case (a). At the onset of an outburst, viscosity increases in the Keplerian disk due to the well-known limit cycle instability. This increases the sub-Keplerian component first because of the smaller infall time (Case b), causing the initial rise of the hard X-ray radiation (CT95), which has been observed in A0620–00 (Ricketts, Pounds and Turner 1975) and GS1124–68 (Miyamoto et al. 1994; Ebisawa et al. 1994; Miyamoto et al. 1995). Subsequently, as the Keplerian component moves in with the viscous time scale, the soft state outburst is produced (Case c). Thus, we believe that our model with two component accretion rates can satisfactorily explain different spectral states of black hole binaries.

## 3  Comparison with Observations

Basically, thermal Comptonization model, in which soft photons are up-scattered by hot thermal plasmas, works for the hard state energy spectra (e.g., Sunyaev and Titarchuk 1980; Titarchuk 1994a). In principle, our low state model can reproduce similar result as in the previous works, but without invoking components such as corona or Compton clouds (e.g., Haardt et al. 1993). In our case, post-shock region acts naturally as the Compton cloud. In what follows, we apply our high state model (where the cooler post-shock flow behaves as a converging inflow) to the LMC X-3 data taken with the GINGA satellite. LMC X-3 is always in the high state, and the hard-tail component is variable independently of the soft component (Treves et al. 1990; Cowley et al. 1992; Ebisawa et al. 1993). We chose the data taken in 1988 January in which the hard-tail component was the strongest. In Figure 2, we show the result of the model fitting. For the soft component, we adopted an optically thick accretion disk model in which local spectrum taking into account electron scattering is analytically solved (Titarchuk 1994b). The hardening factor (color temperature over effective temperature) due to the electron scattering is almost independent of the radius and it is fixed to 1.9 (Shimura & Takahara 1993).

We assumed the distance $d = 50$ kpc and the disk inclination angle is $60°$ (Treves et al. 1988; Kuiper, van Paradijs and van der Klis 1988). The best-fit parameters for the soft component were $M = 5.1 \pm 0.1 M_\odot$ and $\dot{M}_{disk} = 3.2 \pm 0.1 \dot{M}_E$. The disk component explains soft emission at $\lesssim 8$ keV. We fixed the electron temperature of the convergent inflow at 2.5 keV (which is self-consistently attained through the cooling of the post-shock region by disk soft photons, see CT95), and we made $\dot{M}_{conv}$, the net mass accretion rate for the converging inflow and $N_{hard}$, ratio of the disk photons trapped by the converging flow for Comptonization, free parameters. Their best-fit values were $\dot{m} = \dot{M}_{conv}/\dot{M}_E = 4.8 \pm 2.0$ and $N_{hard} = 0.032 \pm 0.015$ respectively. These numbers are very reasonable for the soft-state in our hydrodynamical model. The spectral slope of the hard component is determined by $\dot{M}_{conv}$, and it was $\sim 1.1$ in the present case.

An important prediction of our bulk motion Comptonization model for the soft state is that a power-law spectrum will have a *sharp* high energy cut-off at around the electron rest-mass energy (200 – 500 keV), as opposed to the case of thermal Comptonization (for the hard state) in which a *smooth* turnover takes place with the cut-off at $(1-3)\,kT_e = 50 - 500$ keV depending on the electron temperature. GRS1009-45, a bright transient black hole nova, did not indicate any smooth turnover up to 200keV (Kroeger et al. 1993; Sunyaev et al. 1994), suggesting that the bulk motion Comptonization, as opposed to the thermal Comptonization, may be producing the hard tail. The X-ray Timing Explorer satellite, which will be launched in the end of 1995 or early 1996, is expected to clarify the energy spectra of black hole candidates up to 250 keV with the highest accuracy, and resolve this important issue.

Table 1. Hard Spectral Indices of Black Hole Candidates.

| Source Name | State | Indices | Energy band[a] | Ref. |
|---|---|---|---|---|
| Nova Ophuchi 1977 | soft | 1.4 | 10-200 | [1] |
| EXO1846-031 | soft | 1.4 | 10-25 | [2] |
| GRS1009-45 | soft | 1.53 | 10-100 | [3] |
| GRS1716-249 | hard | 0.5 | 10-100 | [3] |
| GROJ0422+32 | hard | 0.5 | 10-100 | [3] |
| Cyg X-1 | soft | 1.21 | 10-100 | [4] |
| — | hard | 0.85 | 10-300 | [4] |
| GX339-4 | soft | 1.5-1.9 | 10-37 | [5] |
| — | hard | 0.5 | 10-50 | [6] |
| GS1124-68 | soft | 1.6 | 10-37 | [7] |
| — | hard | 0.7 | 10-37 | [7] |
| GS2000+25 | soft | 1.0 | 20-200 | [8] |
| GS2023+338 | hard | 0.3-0.6 | 10-60 | [9] |
| A0620-00 | soft | 1 | 10-200 | [10] |
| LMC X-3 | soft | 1.2 | 10-37 | [11] |
| LMC X-1 | soft | 1.3-1.4[b] | 10-37 | [12] |

References: [1] Wilson & Rothschild (1983); [2] Parmar et al. (1993); [3] Sunyaev et al. (1994); [4] Dolan et al. (1979); [5] Miyamoto et al. (1991); [6] Ricketts (1983); [7] Ebisawa et al. (1994); [8] Dobereiner et al. (1994); [9] Sunyaev et al. (1991); [10] Coe et al. (1976); [11] Ebisawa et al. (1993); [12] Ebisawa et al. (1989)

Notes: a) keV. b) Depends on the choice of the soft-component.

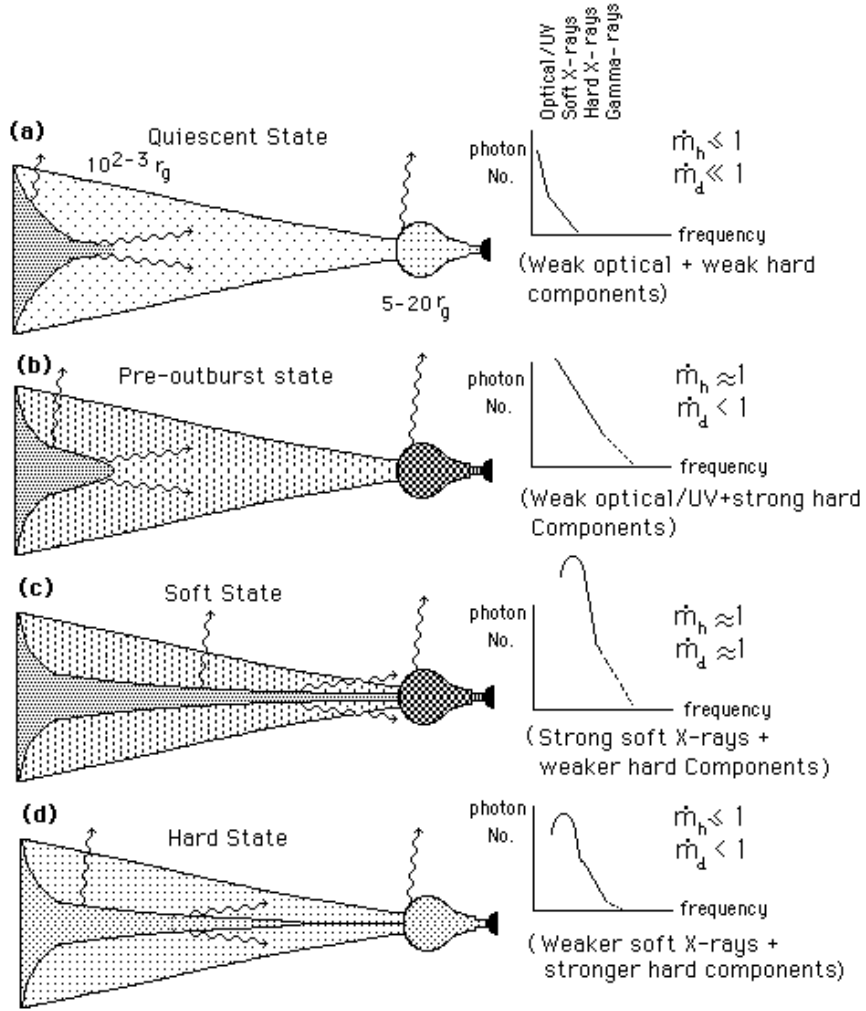

Figure 1: Schematic diagram showing the probable natures of a two-component inflow in roughly logarithmic radial scales. Schematic spectra (solid power-law from thermal Comptonization and the dashed power-law from the bulk motion Comptonization) from each of the cases (a-d) are shown as well. Heavily shaded regions near the left side are the Keplerian disk component, while the lightly shaded regions are the sub-Keplerian components. Hotter, puffed-up accretion shock is close to the hole. The quiescent state (a), pre-outburst state (b), soft state (c) and the hard state (d) are shown.

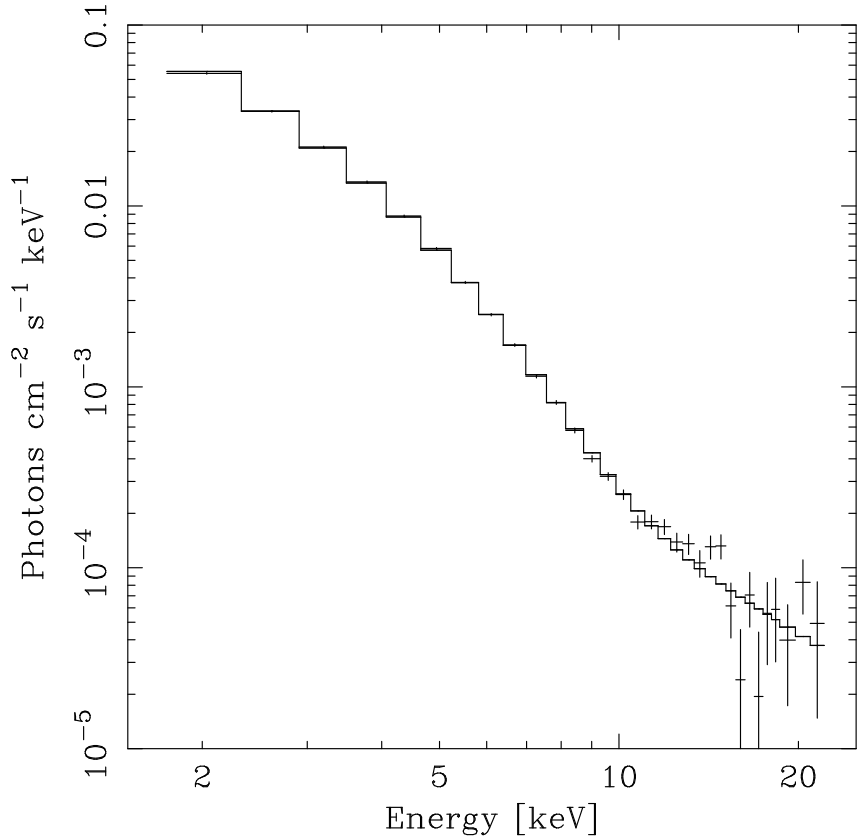

Figure 2: Result of applying our high state model to GINGA LMC X-3 spectrum taken in 1988 January. Detector response is unfolded. Soft component, which is dominant below ∼ 8 keV, is due to an optically thick accretion disk. The hard tail above ∼ 8 keV has a slope of 1.1, and is primarily due to Comptonization of bulk motion in converging inflow.